# Localized Function Method Based on the Galerkin Method Applying a Set of Sine Functions to Model Photonic Crystal Fibres

E. I. Karakoleva[1,*], B. S. Zafirova[1], A. T. Andreev[1]

*Corresponding author: karakoleva@issp.bas.bg
[1] (Institute of Solid State Physics, Bulgarian Academy of Science, Sofia, Bulgaria)

**Abstract:** A development and an application of the localized function method based on the Galerkin method applying a set of Sine functions to approximate the unknown mode fields of the localized modes propagating along the photonic crystal fibres (PCFs) is proposed. A way for considerably reducing the number of integrals in the case of symmetrical holes shapes with respect to the axes of coordinate systems located at the centers of the holes (circular, elliptical etc.) is also presented. The method doesn't require an expansion of the refractive index and thus inaccuracies of the expansion can be avoided. In the case of a circular form of the holes all integrals are solved analytically.

**Keywords:** Photonic crystal fiber; Localized function method; Galerkin method; a set of Sine functions

1. Introduction

Photonic crystal fibres (PCFs) [1] attracted considerable attention since their waveguiding structure provides possibilities for creation of new or improved properties compared to the conventional optical fibres ( for example – endlessly single-mode operation in the entire spectral region of transparency of the fibre's material [2]; unique dispersion properties – ultra flattened dispersion over very wide spectral range, zero dispersion realized at chosen (even visible) wavelengths, multiple zero dispersion [3-5]; very low or very high nonlinearity [6-9]; low or high birefringence [10-12]). Thanks to such properties PCFs find applications to many fields such as: optical communications [13-15], sensor technology [16-18], fibre optic light sources [19, 20], spectroscopy [21, 22], biomedical studies [23, 24], etc.
The high refractive index contrast of the materials requires full-vector methods [25] to model PCFs accurately. The most widely used ones are a plane wave expansion method [26-28], a localized function method (LFM)[29, 30], a beam propagation method (BMP) [31], a finite-element method (FEM) [32, 33], a finite-difference method (FDM) in the time domain [34], a finite-difference method (FDM) in the frequency domain [35-37], a highly accurate semi-analytical multipole method [38, 39]. A brief review of their merits and drawbacks is given in [35].

Here we offer a development and an application of the localized function method based on the Galerkin method applying a set of Sine functions to approximate the unknown mode fields of the localized modes propagating along the PCF. Moreover, since one drawback of the method is the need for a calculation of many integrals [35], we propose a way to reduce considerably their number in the case of symmetrical holes shapes with respect to the axes of the coordinate systems located at the centers of the holes (circular, elliptical etc.). The method doesn't require an expansion of the refractive index unlike PWM and LFM with Hermite-Gaussian functions, which is a major limiting factor for the accuracy of the calculations [25]. Due to this the regions with very small dimensions and a fine structure of interfaces can be considered exactly. In the case of a circular shape of the holes all integrals are solved analytically.

We use some aspects of [40] applied to conventional optic fibres.

## 2. Formulation of the problem

We consider a translationally invariant PCF consisting of $N_h$ holes located in an optical medium (host medium). Monochromatic light with angular frequency ω and time dependence $\exp(i\omega t)$ propagates along PCF in the direction of the axis z. We look for transverse distributions of the modal electric and magnetic fields with respect to a Cartesian coordinate system xOy with an origin at the lower left angle of a rectangular material domain with dimensions $L_x$ (0≤x≤$L_x$) and $L_y$ (0≤y≤$L_y$) comprising the cross-section of PCF with arbitrary locations of the holes. We assume that the electric and magnetic fields are zero at the domain boundaries. With an appropriate choice of the domain dimensions this approximation is reasonable due to the abrupt drop of the fields of the guided modes outside the core.

The modal electric and magnetic fields are solutions of the vector wave equations:

$$\nabla^2 \vec{E} + \nabla\left[\vec{E}.\frac{\nabla n^2}{n^2}\right] + n^2 k^2 \vec{E} = 0 \quad (1) \qquad \nabla^2 \vec{H} + \left[\frac{\nabla n^2}{n^2} \times (\nabla \times \vec{H})\right] + n^2 k^2 \vec{H} = 0 \quad (2)$$

where $\vec{E} \equiv \vec{E}(x,y,z)$ is the electric field vector, $\vec{H} \equiv \vec{H}(x,y,z)$ is the magnetic field vector, $k = \omega(\varepsilon_0 \mu_0)^{1/2}$ is the wave number in free space, $\varepsilon_0$ is the dielectric permittivity of the vacuum, $\mu_0$ is the magnetic permeability of the vacuum and n ≡ n(x,y) is the refractive index of the medium.

In the Cartesian coordinate system the vector wave equations can be decomposed into x, y, and z components. In this decomposition we use the fact that $\partial n/\partial z = 0$. The z-dependence of the components of the fields is assumed to be $\exp(i\beta z)$, where $\beta$ is the longitudinal constant of propagation. It is enough to find solutions for the transverse components of the electric field $E_x(x,y)$, $E_y(x,y)$ and magnetic field $H_x(x,y)$, $H_y(x,y)$, because the longitudinal components $E_z(x,y)$ and $H_z(x,y)$ can be obtained from the transverse ones.

So, we look for solutions of the two systems of two coupled partial differential equations for the unknown x and y components of the electric field $E_x \equiv E_x(x,y), E_y \equiv E_y(x,y)$ and the magnetic field $H_y \equiv H_y(x,y)$, $H_x \equiv H_x(x,y)$:

$$\frac{\partial^2 E_x}{\partial x^2} + \frac{\partial^2 E_x}{\partial y^2} + (n^2 k^2 - \beta^2)E_x + 2\frac{\partial}{\partial x}\left[E_x \frac{\partial \ln(n)}{\partial x} + E_y \frac{\partial \ln(n)}{\partial y}\right] = 0 \quad (3)$$

$$\frac{\partial^2 E_y}{\partial x^2} + \frac{\partial^2 E_y}{\partial y^2} + (n^2 k^2 - \beta^2)E_y + 2\frac{\partial}{\partial y}\left[E_x \frac{\partial \ln(n)}{\partial x} + E_y \frac{\partial \ln(n)}{\partial y}\right] = 0 \quad (4)$$

$$\frac{\partial^2 H_y}{\partial x^2} + \frac{\partial^2 H_y}{\partial y^2} + (n^2 k^2 - \beta^2)H_y - 2\frac{\partial \ln(n)}{\partial x}\left(\frac{\partial H_y}{\partial x} - \frac{\partial H_x}{\partial y}\right) = 0 \quad (5)$$

$$\frac{\partial^2 H_x}{\partial x^2} + \frac{\partial^2 H_x}{\partial y^2} + (n^2 k^2 - \beta^2)H_x + 2\frac{\partial \ln(n)}{\partial y}\left(\frac{\partial H_y}{\partial x} - \frac{\partial H_x}{\partial y}\right) = 0 \quad (6)$$

The equation for $H_x$ is placed after that for $H_y$ because $H_y$ and $H_x$ have a transverse spatial distribution as $E_x$ and $E_y$ for plane waves in free space and also for guided modes. We look for the solutions of these equations in the form:

$$E_x(x,y) = \sum_{\mu=1}^{\infty}\sum_{\nu=1}^{\infty} A_{\mu\nu}^E \Phi_{\mu\nu}(x,y) \quad (7) \qquad E_y(x,y) = \sum_{\mu=1}^{\infty}\sum_{\nu=1}^{\infty} B_{\mu\nu}^E \Phi_{\mu\nu}(x,y) \quad (8)$$

$$H_y(x,y) = \sum_{\mu=1}^{\infty}\sum_{\nu=1}^{\infty} A_{\mu\nu}^H \Phi_{\mu\nu}(x,y) \quad (9) \qquad H_x(x,y) = \sum_{\mu=1}^{\infty}\sum_{\nu=1}^{\infty} B_{\mu\nu}^H \Phi_{\mu\nu}(x,y) \quad (10)$$

where: $\Phi_{\mu\nu}(x,y) = [2/(L_x L_y)^{1/2}]\sin(\sigma_\mu x)\sin(\rho_\nu y)$ (11)

is a complete orthonormal set of sine functions which are orthogonal over the finite rectangular domain with dimensions $L_x$ ($0 \le x \le L_x$) and $L_y$ ($0 \le y \le L_y$):

$$\int_0^{L_x} dx \int_0^{L_y} dy \, \Phi_{\mu\nu}(x,y) \Phi_{\mu'\nu'}(x,y) = \delta_{\mu\mu'}\delta_{\nu\nu'}, \quad (12)$$

$\sigma_\mu = (\mu\pi/L_x)$; $\rho_\nu = (\nu\pi/L_y)$; $\mu, \nu$ are integers; $A^E_{\mu\nu}, B^E_{\mu\nu}, A^H_{\mu\nu}, B^H_{\mu\nu}$ are unknown coefficients in the expansions of $E_x$, $E_y$, $H_y$ and $H_x$, respectively.

Using the Galerkin method each member of the systems of partial differential equations (3), (4) and (5), (6) is multiplied with $\Phi_{\mu'\nu'}$ and the equations are integrated over the surface $S = L_x L_y$ of the material domain. $E_x$, $E_y$, $H_y$ and $H_x$ are replaced by their expansions (7), (8), (9) and (10). The differentiations are performed and the orthogonality of the sine functions is used. The derivatives of $\ln(n)$ produce delta-functions at the interfaces. They can be removed by integrating by parts. The residual terms produced by this method vanish because $\Phi_{\mu\nu}$ are zero at the boundaries of the domain S.

In this way the two systems of the two partial differential equations are converted into two systems each of $2m_x m_y$ coupled linear algebraic equations ($m_x$ and $m_y$ are the numbers of members in truncated sums) for the unknown coefficients $A^E_{\mu\nu}, B^E_{\mu\nu}, A^H_{\mu\nu}, B^H_{\mu\nu}$:

$$\sum_{\mu=1}^{m_x}\sum_{\nu=1}^{m_y}(M^E_{\mu'\nu',\mu\nu}A^E_{\mu\nu} + N^E_{\mu'\nu',\mu\nu}B^E_{\mu\nu}) = (\beta/k)^2 A^E_{\mu'\nu'} \quad (13)$$

$$\sum_{\mu=1}^{m_x}\sum_{\nu=1}^{m_y}(R^E_{\mu'\nu',\mu\nu}A^E_{\mu\nu} + S^E_{\mu'\nu',\mu\nu}B^E_{\mu\nu}) = (\beta/k)^2 B^E_{\mu'\nu'} \quad \mu' = 1,2,...,m_x, \quad \nu' = 1,2,...,m_y \quad (14)$$

$$\sum_{\mu=1}^{m_x}\sum_{\nu=1}^{m_y}(M^H_{\mu'\nu',\mu\nu}A^H_{\mu\nu} + N^H_{\mu'\nu',\mu\nu}B^H_{\mu\nu}) = (\beta/k)^2 A^H_{\mu'\nu'} \quad (15)$$

$$\sum_{\mu=1}^{m_x}\sum_{\nu=1}^{m_y}(R^H_{\mu'\nu',\mu\nu}A^H_{\mu\nu} + S^H_{\mu'\nu',\mu\nu}B^H_{\mu\nu}) = (\beta/k)^2 B^H_{\mu'\nu'} \quad \mu' = 1,2,...,m_x, \quad \nu' = 1,2,...,m_y \quad (16)$$

where:
$$M^E_{\mu'\nu',\mu\nu} = \frac{4}{S}\int_0^{L_x} dx \int_0^{L_y} dy \left[(n^2 - n^2_{\mu\nu})P_{ssss} + 2\frac{\sigma_{\mu'}}{k^2}\ln(n)(\sigma_\mu P_{ccss} - \sigma_{\mu'} P_{ssss})\right] \quad (17)$$

$$N^E_{\mu'\nu',\mu\nu} = \frac{8}{S}\frac{\sigma_{\mu'}}{k^2}\int_0^{L_x} dx \int_0^{L_y} dy \ln(n)(\rho_\nu P_{sccs} + \rho_{\nu'} P_{scsc}) \quad (18)$$

$$R^E_{\mu'\nu',\mu\nu} = \frac{8}{S}\frac{\rho_{\nu'}}{k^2}\int_0^{L_x} dx \int_0^{L_y} dy \ln(n)(\sigma_\mu P_{cssc} + \sigma_{\mu'} P_{scsc}) \quad (19)$$

$$S^E_{\mu'v',\mu v} = \frac{4}{S}\int_0^{L_x} dx \int_0^{L_y} dy \left[ (n^2 - n^2_{\mu v})P_{ssss} + 2\frac{\rho_{v'}}{k^2}\ln(n)(\rho_v P_{sscc} - \rho_{v'} P_{ssss}) \right] \quad (20)$$

$$M^H_{\mu'v',\mu v} = \frac{4}{S}\int_0^{L_x} dx \int_0^{L_y} dy \left[ (n^2 - n^2_{\mu v})P_{ssss} + 2\frac{\sigma_\mu}{k^2}\ln(n)(\sigma_{\mu'} P_{ccss} - \sigma_\mu P_{ssss}) \right] \quad (21)$$

$$N^H_{\mu'v',\mu v} = -\frac{8}{S}\frac{\rho_v}{k^2}\int_0^{L_x} dx \int_0^{L_y} dy \ln(n)(\sigma_\mu P_{cscs} + \sigma_{\mu'} P_{sccs}) \quad (22)$$

$$R^H_{\mu'v',\mu v} = -\frac{8}{S}\frac{\sigma_\mu}{k^2}\int_0^{L_x} dx \int_0^{L_y} dy \ln(n)(\rho_v P_{cscs} + \rho_{v'} P_{cssc}) \quad (23)$$

$$S^H_{\mu'v',\mu v} = \frac{4}{S}\int_0^{L_x} dx \int_0^{L_y} dy \left[ (n^2 - n^2_{\mu v})P_{ssss} + 2\frac{\rho_v}{k^2}\ln(n)(\rho_{v'} P_{sscc} - \rho_v P_{ssss}) \right] \quad (24)$$

$$n \equiv n(x,y) = \begin{cases} n_i = const, & x,y \in S_i \\ n_{host} = const, & x,y \notin S_i \end{cases}, \quad i = 1,2,...,N_h$$

$n_i$ is the constant refractive index of the ith hole with the surface $S_i$, $n_{host}$ is the constant refractive index of the host medium, $N_h$ is the number of the holes into the photonic crystal fibre, $n^2_{\mu v} \equiv (\sigma^2_\mu + \rho^2_v)/k^2$ is a dimensionless quantity,

$P_{ssss} \equiv P_{ssss}(x,y) = \sin(\sigma_\mu x)\sin(\sigma_{\mu'} x)\sin(\rho_v y)\sin(\rho_{v'} y)$

$P_{ccss} \equiv P_{ccss}(x,y) = \cos(\sigma_\mu x)\cos(\sigma_{\mu'} x)\sin(\rho_v y)\sin(\rho_{v'} y)$.

The definitions of the remaining products are analogous.

Then each of the integrals in (17) – (24) can be presented as a sum of double integrals over the host medium and over the holes in it:

$$\int_0^{L_x} \int_0^{L_y} f(x,y;n(x,y)) dx dy = \sum_{i=1}^{N_h} \iint_{S_i} f(x,y;n_i) dx dy + \iint_{host\ medium} f(x,y;n_{host}) dx dy \quad (25)$$

In order not to integrate over the host medium we add and subtract integrals over the holes surfaces, in which the refractive indices are replaced by the refractive index of the host medium and obtain:

$$\int_0^{L_x} \int_0^{L_y} f(x,y;n(x,y)) dx dy = \int_0^{L_x} \int_0^{L_y} f(x,y;n_{host}) dx dy + \sum_{i=1}^{N_h} \iint_{S_i} \left[ f(x,y;n_i) - f(x,y;n_{host}) \right] dx dy \quad (26)$$

e.g. the double integral over the domain with $N_h$ interfaces is replaced by a sum of a double integral over a homogeneous medium with a surface S and refractive index $n_{host}$ (where the orthogonality of the sine functions can be used) and $N_h$ homogeneous media with surfaces $S_i$, i=1, 2,…, $N_h$ with changed refractive indices.

Then the expressions (17) – (24) can be written as:

$$M^E_{\mu'\nu',\mu\nu} = \frac{4}{S} \sum_{i=1}^{N_h+1} \left[ n^i_s I^i_{ssss} + 2\frac{\sigma_{\mu'}}{k^2} \ln(n^i_d)\left(\sigma_\mu I^i_{ccss} - \sigma_{\mu'} I^i_{ssss}\right) \right]$$

$$N^E_{\mu'\nu',\mu\nu} = \frac{8}{S} \sum_{i=1}^{N_h+1} \frac{\sigma_{\mu'}}{k^2} \ln(n^i_d)\left(\rho_\nu I^i_{sccs} + \rho_{\nu'} I^i_{scsc}\right)$$

$$R^E_{\mu'\nu',\mu\nu} = \frac{8}{S} \sum_{i=1}^{N_h+1} \frac{\rho_{\nu'}}{k^2} \ln(n^i_d)\left(\sigma_\mu I^i_{cssc} + \sigma_{\mu'} I^i_{scsc}\right)$$

$$S^E_{\mu'\nu',\mu\nu} = \frac{4}{S} \sum_{i=1}^{N_h+1} \left[ n^i_s I^i_{ssss} + 2\frac{\rho_{\nu'}}{k^2} \ln(n^i_d)\left(\rho_\nu I^i_{sscc} - \rho_{\nu'} I^i_{ssss}\right) \right]$$

$$M^H_{\mu'\nu',\mu\nu} = \frac{4}{S} \sum_{i=1}^{N_h+1} \left[ n^i_s I^i_{ssss} + 2\frac{\sigma_\mu}{k^2} \ln(n^i_d)\left(\sigma_{\mu'} I^i_{ccss} - \sigma_\mu I^i_{ssss}\right) \right]$$

$$N^H_{\mu'\nu',\mu\nu} = -\frac{8}{S} \sum_{i=1}^{N_h+1} \frac{\rho_\nu}{k^2} \ln(n^i_d)\left(\sigma_\mu I^i_{cscs} + \sigma_{\mu'} I^i_{sccs}\right)$$

$$R^H_{\mu'\nu',\mu\nu} = -\frac{8}{S} \sum_{i=1}^{N_h+1} \frac{\sigma_\mu}{k^2} \ln(n^i_d)\left(\rho_\nu I^i_{cscs} + \rho_{\nu'} I^i_{cssc}\right)$$

$$S^H_{\mu'\nu',\mu\nu} = \frac{4}{S} \sum_{i=1}^{N_h+1} \left[ n^i_s I^i_{ssss} + 2\frac{\rho_\nu}{k^2} \ln(n^i_d)\left(\rho_{\nu'} I^i_{sscc} - \rho_\nu I^i_{ssss}\right) \right]$$

where:

$$n^i_s = \begin{cases} n_i^2 - n_{host}^2, & i = 1,2,...,N_h \\ n_{host}^2 - n_{\mu\nu}^2, & i = N_h + 1 \end{cases} ; \quad n^i_d = \begin{cases} n_i / n_{host}, & i = 1,2,...,N_h \\ n_{host}, & i = N_h + 1 \end{cases} ;$$

$$I^i_{ssss} = \begin{cases} I^{S_i}_{ssss} = \iint_{S_i} P_{ssss}(x,y)\,dxdy & i = 1,2,...,N_h \\ I^S_{ssss} = \iint_S P_{ssss}(x,y)\,dxdy & i = N_h + 1 \end{cases}$$

The number $N_h + 1$ is referred to the material domain. The definitions of the remaining integrals are analogous.

Let us consider the ith hole. A local coordinate system $x'O_i y'$ with an origin located at the center of the ith hole and axes parallel to the axes of the global coordinate system xOy is introduced (Fig. 1):

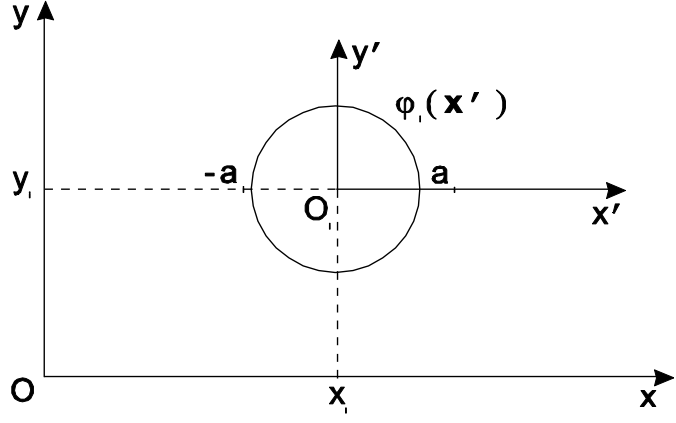

**Fig.1.** A local coordinate system $x'O_iy'$ with an origin located at the center of the ith hole of a PCF and axes parallel to the axes of the global coordinate system xOy.

We consider the case when the hole form is symmetrical with respect to the local axes. After variable changing the integral $I^{S_i}_{ssss}$ can be written in the local coordinate system as:

$$I^{S_i}_{ssss} = \int_{-a_i}^{a_i} dx' \sin[\sigma_\mu(x_i + x')]\sin[\sigma_{\mu'}(x_i + x')] \int_{-\varphi_i(x')}^{\varphi_i(x')} dy' \sin[\rho_\nu(y_i + y')]\sin[\rho_{\nu'}(y_i + y')] =$$

$$= \left\{ \iint dx'dy' \cos[\sigma_m(x_i+x')]\cos[\rho_m(y_i+y')] - \iint dx'dy' \cos[\sigma_m(x_i+x')]\cos[\rho_p(y_i+y')] - \right.$$

$$\left. - \iint dx'dy' \cos[\sigma_p(x_i+x')]\cos[\rho_m(y_i+y')] + \iint dx'dy' \cos[\sigma_p(x_i+x')]\cos[\rho_p(y_i+y')] \right\}/4$$

where $x_i$ and $y_i$ are coordinates of the center of the ith hole in the global coordinate system; $-a_i$ and $a_i$ are lower and upper limit of changing of the local variable $x'$; $-\varphi_i(x')$, $\varphi_i(x')$ are lower and upper limit of changing of the local variable $y'$ (the integration limits are dropped for brevity in the last expression); $\sigma_m \equiv \sigma_\mu - \sigma_{\mu'}$; $\sigma_p \equiv \sigma_\mu + \sigma_{\mu'}$; $\rho_m \equiv \rho_\nu - \rho_{\nu'}$; $\rho_p \equiv \rho_\nu - \rho_{\nu'}$.

The global variables $x_i$ and $y_i$ can be separated from the local variables $x'$ and $y'$ using the formulae for adding of trigonometric functions and the fact that the integral over odd function is equal to zero. Then:

$$\iint dx'dy' \cos[\sigma_j(x_i+x')]\cos[\rho_k(y_i+y')] =$$

$$= \cos(\sigma_j x_i)\cos(\rho_k y_i)\iint dx'dy' \cos(\sigma_j x')\cos(\rho_k y'),\ j,k = m,p.$$

This presentation makes it possible to reduce the number of integrals to the four integrals $I_{mm}$, $I_{mp}$, $I_{pm}$, $I_{pp}$:

$$I^{S_i}_{ssss} = \left( c_{mm}I_{mm} - c_{mp}I_{mp} - c_{pm}I_{pm} + c_{pp}I_{pp} \right)/4$$

$$I^{S_i}_{ccss} = \left( c_{mm}I_{mm} - c_{mp}I_{mp} + c_{pm}I_{pm} - c_{pp}I_{pp} \right)/4$$

$$I^{S_i}_{sscc} = \left( c_{mm}I_{mm} + c_{mp}I_{mp} - c_{pm}I_{pm} - c_{pp}I_{pp} \right)/4$$

$$I^{S_i}_{sccs} = \left( -s_{mm}I_{mm} + s_{mp}I_{mp} - s_{pm}I_{pm} + s_{pp}I_{pp} \right)/4$$

$$I^{S_i}_{scsc} = \left( s_{mm}I_{mm} + s_{mp}I_{mp} + s_{pm}I_{pm} + s_{pp}I_{pp} \right)/4$$

$$I^{S_i}_{cssc} = \left( -s_{mm}I_{mm} - s_{mp}I_{mp} + s_{pm}I_{pm} + s_{pp}I_{pp} \right)/4$$

$$I^{S_i}_{cscs} = \left( s_{mm}I_{mm} - s_{mp}I_{mp} - s_{pm}I_{pm} + s_{pp}I_{pp} \right)/4$$

where $c_{jk} = \cos(\sigma_j x_i)\cos(\rho_k y_i)$, $s_{jk} = \sin(\sigma_j x_i)\sin(\rho_k y_i)$ are functions depending only on the location of the holes with respect to the global coordinate system;

$$I_{jk} = \int_{-a_i}^{a_i} \int_{-\varphi_i(x')}^{\varphi_i(x')} dx'dy' \cos(\sigma_j x')\cos(\rho_k y') \qquad j,k = m, p$$ are integrals which depend on the hole shape, but not on its location. For all holes with identical shapes the four integrals can be solved only once. This reduces considerably the integrals numbers and the calculation time especially when the integrals are numerically calculated. These four integrals can be solved analytically when the hole cross section has a circular form. In this case:

$$I_{ssss}^{S_i} = \int_{x_i-a_i}^{x_i+a_i} dx \sin(\sigma_\mu x)\sin(\sigma_{\mu'} x) \int_{y_i-\sqrt{a_i^2-(x-x_i)^2}}^{y_i+\sqrt{a_i^2-(x-x_i)^2}} dy \; \sin(\rho_\nu y)\sin(\rho_{\nu'} y) =$$

$$= \int_{-a_i}^{a_i} dx' \sin[\sigma_\mu(x_i+x')]\sin[\sigma_{\mu'}(x_i+x')] \int_{-\sqrt{a_i^2-(x')^2}}^{\sqrt{a_i^2-(x')^2}} dy' \; \sin[\rho_\nu(y_i+y')]\sin[\rho_{\nu'}(y_i+y')] =$$

$$= (c_{mm}I_{mm} - c_{mp}I_{mp} - c_{pm}I_{pm} + c_{pp}I_{pp})/4$$

where $I_{jk} = \int_{-a_i}^{a_i} \int_{-\sqrt{a_i^2-(x')^2}}^{\sqrt{a_i^2-(x')^2}} dx'dy' \cos(\sigma_j x')\cos(\rho_k y')$, $j,k = m, p$. In order to solve analytically the double integral the following identity [40] is used:

$$\int_{-a}^{a} dx \int_{-\sqrt{a^2-x^2}}^{\sqrt{a^2-x^2}} dy \cos(Ax + By) = \frac{2\pi a}{\sqrt{A^2+B^2}} J_1\left(a\sqrt{A^2+B^2}\right),$$ where the used here coordinates x, y and parameters A, B and a are related only to this identity and are not connected with the global and local variables and parameters used so far. $J_1$ is the Bessel function of order 1 [41]. The integral $\iint dx'dy' \sin(\sigma_j x')\sin(\rho_k y')$ is subtracted from $I_{jk}$ and we obtain:

$$I_{jk} = \int_{-a_i}^{a_i} \int_{-\sqrt{a_i^2-(x')^2}}^{\sqrt{a_i^2-(x')^2}} dx'dy' \cos(\sigma_j x' + \rho_k y') = \frac{2\pi a_i}{\sqrt{\sigma_j^2+\rho_k^2}} J_1\left(a_i\sqrt{\sigma_j^2+\rho_k^2}\right) \quad j,k = m, p.$$

So, in the case of circular holes all integrals are solved analytically.

The systems of algebraic equations (13), (14) and (15), (16) can be written in the form of matrix eigenvalue equations: $\hat{C}^E \vec{X}^E = (\beta/k)^2 \vec{X}^E$; $\hat{C}^H \vec{X}^H = (\beta/k)^2 \vec{X}^H$, where

$$\hat{C}^E \equiv \begin{pmatrix} \hat{M}^E & \hat{N}^E \\ \hat{R}^E & \hat{S}^E \end{pmatrix}; \quad \hat{C}^H \equiv \begin{pmatrix} \hat{M}^H & \hat{N}^H \\ \hat{R}^H & \hat{S}^H \end{pmatrix} \quad ; \hat{M}^E, \hat{N}^E, \hat{R}^E, \hat{S}^E, \hat{M}^H, \hat{N}^H, \hat{R}^H, \hat{S}^H$$

are matrices consisting of the coefficients $M_{\mu'\nu',\mu\nu}^E; N_{\mu'\nu',\mu\nu}^E; R_{\mu'\nu',\mu\nu}^E; S_{\mu'\nu',\mu\nu}^E$; $M_{\mu'\nu',\mu\nu}^H; N_{\mu'\nu',\mu\nu}^H; R_{\mu'\nu',\mu\nu}^H; S_{\mu'\nu',\mu\nu}^H$, respectively; $\vec{X}^E = (\vec{A}^E, \vec{B}^E)^T$ and $\vec{X}^H = (\vec{A}^H, \vec{B}^H)^T$ are eigenvectors, consisting of coefficients $A_{\mu\nu}^E; B_{\mu\nu}^E; A_{\mu\nu}^H; B_{\mu\nu}^H$ and $(\beta/k)^2$ are unknown eigenvalues.

The above derivation is incorporated into a single Visual Fortran 6.5 code. It calculates matrices elements, modal effective indices and transverse components of both the electric and magnetic field propagating along the PCF. EISPACK [42] is incorporated into the code and is used to solve the eigenvalue equations.

## 3. Numerical results

The PCF example calculated by the multipole method is used in this paper since here analytical results are available for comparison. The PCF consists of 6 cylindrical air holes each with a diameter 1.0μm and refractive index $n_i$ =1.0 (i=1, 2,..., 6) arranged in a hexagonal lattice with a constant (a pitch) Λ=2.3μm within a host medium with refractive index $n_{host}$=1.44390356 (Fig. 2). According to the multipole method [43, 44], the accurate effective index of the fundamental mode at wavelength λ=1.56μm would be $1.42078454+i7.20952 \times 10^{-4}$.

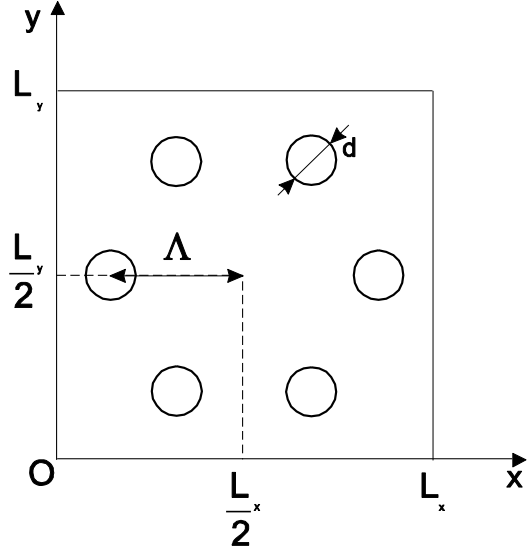

**Fig. 2** Cross-section of a PCF consisting of 6 identical cylindrical air holes with a diameter d=1.0μm arranged in a hexagonal lattice with a constant (a pitch) Λ=2.3μm.

Here we are solving the eigenvalue equation for a transverse magnetic field which is continuous at the interfaces unlike the electric field and requires less sine functions to be approximated accurately. In order to find the solution the numbers of expansion terms and the dimensions of the material domain are changed from 5 to 71 and from 2Λ+0.2 [μm] to 4Λ, respectively. The essential part of the results is shown in the Table 1. It can be seen from the Table that the value of $n_{eff}$ with smallest relation errors $\Delta_m$=-9.520x$10^{-7}$ and $\Delta_L$=2.400x$10^{-8}$ is a solution of the problem, $n_{eff}$ =1.420790647751640. The radiation loss is neglected. The relative error between this solution and the one obtained by a multipole method is -4.3x$10^{-6}$.

**Table 1.** The values of the effective index for the fundamental mode. Here $n_{eff}$ are solutions of the effective index for which the smallest relative error $\Delta_m$ between two successive solutions with different terms in their expansions is obtained at dimensions $L_x$ and $L_y$. The values of $m_x$ and $m_y$ are the expansion terms necessary to achieve $\Delta_m$. It is assumed that $m_x$=$m_y$ and $L_x$=$L_y$. $\Delta_L$ are the relative errors for two solutions at two successive values of the dimensions of the material domain.

| $L_x$=$L_y$ [μm] | $m_x$=$m_y$ | neff | $\Delta_m$ | $\Delta_L$ |
|---|---|---|---|---|
| 6.80 | 70 | 1.420719731100830 | -1.410 x 10-6 | -- |
| 6.82 | 70 | 1.420728725549000 | -1.410 x 10-6 | 6.330 x 10-6 |
| 6.84 | 68 | 1.420737677303520 | +1.010 x 10-7 | 6.300 x 10-6 |
| 6.86 | 64 | 1.420747668278230 | +1.800 x 10-7 | 7.032 x 10-6 |
| 6.88 | 70 | 1.420755232661020 | -1.421 x 10-6 | 5.324 x 10-6 |
| 6.90 | 70 | 1.420763935483380 | -8.610 x 10-6 | 6.120 x 10-6 |
| 6.92 | 66 | 1.420775176219890 | -3.780 x 10-7 | 7.911 x 10-6 |
| 6.94 | 60 | 1.420790612680160 | -2.490 x 10-6 | 10.865 x 10-6 |
| 6.96 | 68 | 1.420790647751640 | -9.520 x 10-7 | 2.400 x 10-8 |
| 6.98 | 70 | 1.420798387400970 | -9.940 x 10-7 | 5.446 x 10-6 |

The vector distribution of the transverse magnetic field $\vec{H}_t(x,y)$ over the PCF cross section and the contour map of the solution are given in Fig.3 and Fig.4.

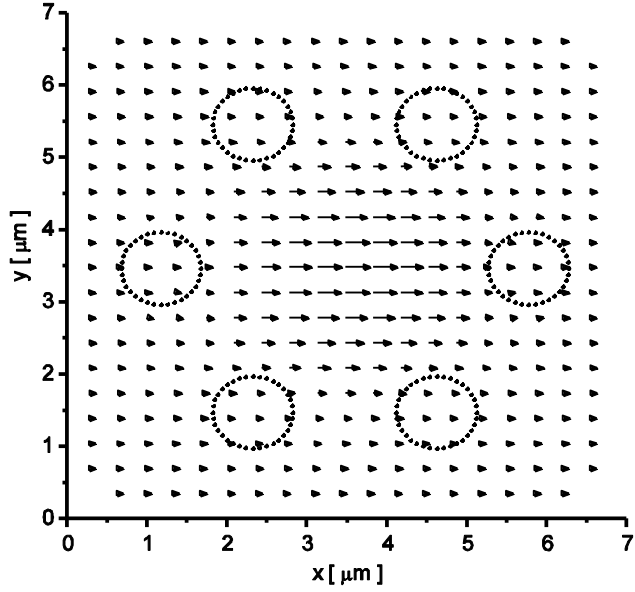

**Fig.3** A vector distribution of the transverse magnetic field $\vec{H}_t(x,y)$ of the fundamental mode over the cross section of the PCF. The refractive index of the air holes is $n_i=1.0$ (i=1, 2,... 6). The effective index is $n_{eff}$=1.420790647751640, $m_x=m_y=68$ are expansion terms, $L_x=L_y=6.96\mu m$ are dimensions of the material domain.

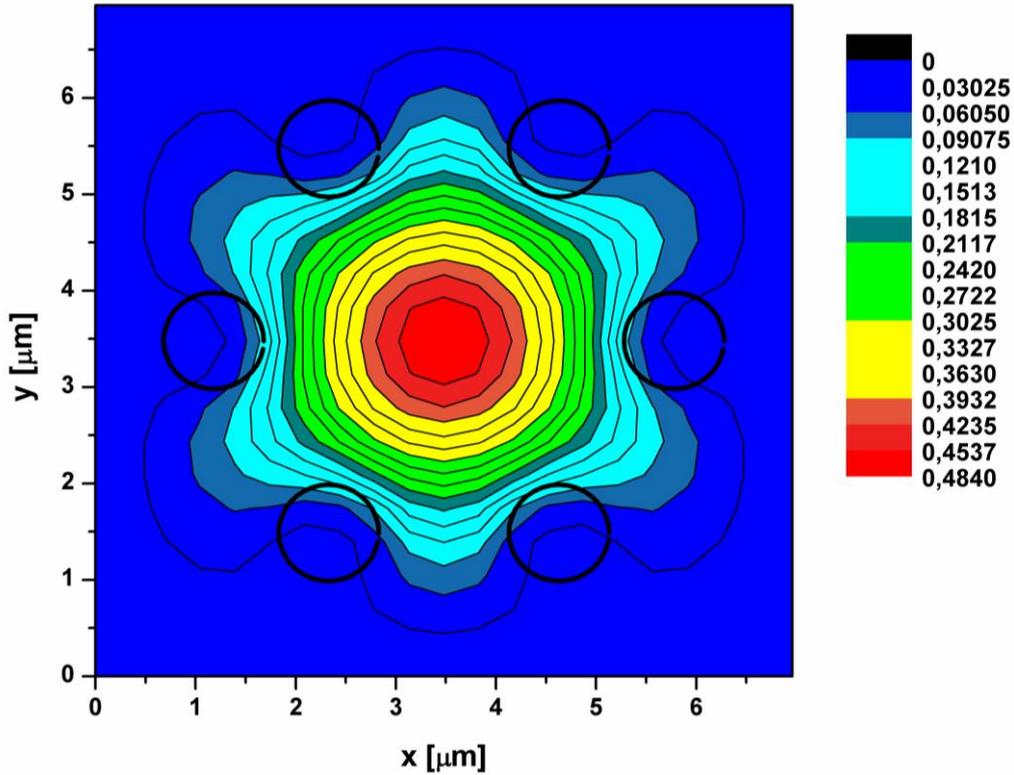

**Fig. 4** A contour map of the magnetic field of the fundamental mode of the PCF (shown in Fig.2), linearly polarized in the positive direction of the axis x. The effective index is $n_{eff}$ =1.420790647751640, $m_x=m_y=68$ are expansion terms, $L_x=L_y=6.96\mu m$ are dimensions of the material domain ($0\leq x \leq L_x$, $0\leq y \leq L_y$).

## 4. Conclusions

We have presented a development and an application of the localized function method based on Galerkin method applying a set of Sine functions for numerical modeling of PCFs with arbitrary locations of holes without expanding of the refractive index. The above derivation is realized as a single Visual Fortran 6.5 code. It calculates matrix elements, modal effective indices and transverse components of both the electric and magnetic fields propagating along the PCF. EISPACK [42] is incorporated in the code and is used to solve the eigenvalue equations. Numerical results obtained by the code are presented for the case of circular holes. The code has possibility to calculate modal effective indices and transverse components of electric and magnetic fields for a PCF with arbitrary locations of holes with square and rectangular shapes (in addition the axes of local coordinate systems at the centers of the holes are rotated with respect to the axes of the global one).The number of the integrals is reduced considerably in the case of symmetric holes shapes. For every element of matrices $\hat{M}, \hat{N}, \hat{R}, \hat{S}$ for both the electric and magnetic fields $7N_h$ integrals are solved for the holes. In the case of identical symmetrical shapes their number is 4.The double integrals for circular, square and rectangular holes are calculated analytically as well as these related to the host medium. In the test example the number of terms in the expansion of the magnetic field over Sine functions is relatively small, about 70 terms along each of axes.